# Optimal control for maximally creating and maintaining a superposition state of a two-level system under the influence of Markovian decoherence

Yukiyoshi Ohtsuki*, Suicho Mikami, and Toru Ajiki

Department of Chemistry, Graduate School of Science, Tohoku University

6-3 Aramaki Aza-Aoba, Aoba-ku, Sendai 980-8578, Japan

David J. Tannor

Department of Chemical Physics, Weizmann Institute of Science, 76100 Rehovot, Israel

* Corresponding author: yukiyoshi.ohtsuki.d2@tohoku.ac.jp

**Abstract**

Reducing decoherence is an essential step toward realizing general-purpose quantum computers beyond the present noisy intermediate-scale quantum (NISQ) computers. To this end, dynamical decoupling (DD) approaches in which external fields are applied to qubits are often adopted. We numerically study DD using a two-level model system (qubit) under the influence of Markovian decoherence by using quantum optimal control theory with slightly modified settings, in which the physical objective is to maximally create and maintain a specified superposition state in a specified control period. An optimal pulse is numerically designed while systematically varying the values of dephasing, population decay, pulse fluence, and control period as well as using two kinds of objective functionals. Although the decrease in purity due to the decoherence gives rise to the upper limit of the target expectation value, i.e., the saturated value, the optimally shaped pulse effectively deals with the decoherence by gradually creating the target superposition state to realize the saturated value as much as possible.

## I. INTRODUCTION

The declaration of quantum computational supremacy by Google scientists [1, 2] has significantly impacted theoretical chemists because one of the promising applications of quantum computers is believed to be chemistry [2-5]. For example, quantum computers can perform full-CI electronic state calculations by using phase estimation algorithms [4]. Unfortunately, the limited number of qubits of present-day quantum computers has prevented the realization of fault-tolerant computation by quantum error correction, that is, general-purpose quantum computation. The current state of quantum computers is often referred to as the era of noisy intermediate-scale quantum (NISQ) machines [6]. In this regard, it would be natural to consider the combined use of classic computers and NISQ computers, for instance, by adopting the variational quantum eigensolver (VQE) algorithm [7]. Examples include the potential curves of hydrogen chains and diazene isomerization [8], and those of water along with several coordinates [9]. Another approach called error mitigation [10, 11] has been designed to cooperate with noise (error) and applied to hydrogen and lithium hydride molecules [12].

To go beyond the NISQ computers, it is essential to reduce decoherence, namely, the system coherence leakage to its environment due to quantum entanglement. Dynamical decoupling (DD) techniques are often used for this purpose [13-21]. The widely employed Carr-Purcell-Meiboom-Gill (CPMG) technique [13, 14] applies an evenly spaced π pulse sequence to suppress phase error accumulation. Such DD techniques as the Uhrig DD [16] and the optimized DD [17] can filter out a certain range of the noise spectrum. In addition to the pulse sequence approaches, there exist DD techniques that utilize continuous fields to remove the specified coupling between the qubits and the environment [20, 21].

The purpose of the present study is to numerically examine the optimized DD pulse sequence. We assume that the qubit, i.e., the two-level system, is initially in the ground state as assumed in the widely used CPMG technique. Under slightly modified settings, we adopt the optimal control theory [22-24], which has undergone rigorous development for the coherent control of molecular dynamics, to maximally create and maintain a specified superposition state under the influence of Markovian decoherence. Because of the simple model, we can numerically design the optimal pulses while systematically varying the values of dephasing, population decay, pulse fluence, and control period. We also discuss the dependence of the optimal pulses on control objective quantification. In addition to its own interest, the numerical results will serve as a reference for more general and advanced optimal DD investigations. This paper is organized as follows. In Sec. II, we outline the optimal control methods used in the present study. In Sec. III, the numerical results are shown after briefly explaining the numerical details. We summarize the present study in Section IV.

## II. Theory

We consider a two-level model system specified by the state $|0\rangle$ with the energy eigenvalue $\hbar\omega_0$ and the state $|1\rangle$ with $\hbar\omega_1$, where $\omega_{10} = \omega_1 - \omega_0 > 0$ is assumed. Through the (electric or magnetic) dipole moment operator $\mu$, the system interacts with an external-field pulse, which will be simply called a pulse in the present study. For notational convenience, the pulse is denoted as $E(t)$ although it can be a magnetic field pulse. The system Hamiltonian is expressed as

$$H(t) = H_0 + V(t) = H_0 - \mu E(t). \tag{1}$$

We assume that the system is surrounded by an environment that does not directly interact with the pulse [25, 26]. The decoherence induced by the environment is described by the relaxation operator $\Gamma$ with phenomenological parameters and is assumed to induce the dephasing and population decay of the system. The equation of motion for the system (reduced) density operator $\rho(t)$ is expressed as

$$\frac{\partial}{\partial t}|\rho(t)\rangle\rangle = -\frac{i}{\hbar}L(t)|\rho(t)\rangle\rangle - \Gamma|\rho(t)\rangle\rangle \equiv -\frac{i}{\hbar}L_{\text{eff}}(t)|\rho(t)\rangle\rangle \qquad (2)$$

in the Liouville-space notation [26]. Here, the Liouville-space operators, $L(t)$, $L_0$, and $M$, correspond to the commutators, $[H(t),\cdots]$, $[H_0,\cdots]$, and $[\mu,\cdots]$, respectively.

Our aim is to find an optimal pulse that maximally creates and maintains a specified superposition state during a specified control period. To quantify our physical objective, we introduce a target operator $W(t)$ that measures the degree of control achievement. We consider two kinds of objective functionals to define the optimal control problems. One of them, referred to as the type I functional, evaluates the degree of control achievement as the integral of the expectation value of $W(t)$, which will be called the target expectation value in the following. The type I functional is expressed as

$$F_I = 2\operatorname{Re}\int_0^{t_f} dt\, y(t)\langle\langle W(t)|\rho(t)\rangle\rangle = 2\operatorname{Re}\int_0^{t_f} dt\, y(t)\operatorname{Tr}\{W^\dagger(t)\rho(t)\}, \qquad (3)$$

where the envelope function $y(t)$ specifies the control period. The second equal sign in Eq. (3) shows the definition of the inner product between the Liouville space vectors. The other functional, which is referred to as the type II functional, is expressed as

$$F_{II} = \int_0^{t_f} dt\, \left|\langle\langle W(t)|\rho(t)\rangle\rangle\right|^2 y(t). \qquad (4)$$

We also introduce the so-called penalty term due to the pulse fluence [27, 28]

$$P = \int_0^{t_f} dt\, \frac{[E(t)]^2}{\hbar A(t)}, \qquad (5)$$

where the positive function $A(t)$ evaluates the physical significance of the penalty. From Eqs. (3) to (5), our optimal control problems end up with the maximization of $J_I = F_I - P$ and $J_{II} = F_{II} - P$. When we maximize the functional without the penalty term $P$, we will refer to the problem by the penalty-free optimization.

The optimal pulses are obtained by solving the coupled pulse-design equations, which are derived by applying the calculus of variations to the maximization problems subject to the constraints due to the equation of motion in Eq. (2). The first-order variation of the density operator is expressed as

$$|\delta\rho(t)\rangle\rangle = \frac{i}{\hbar}\int_0^t G(t,\tau) M |\rho(\tau)\rangle\rangle \delta E(\tau), \tag{6}$$

where the Liouville-space time evolution operator is defined by

$$G(t,\tau) = \hat{T}\exp\left[\frac{i}{\hbar}\int_\tau^t L_{\text{eff}}(\tau_1)d\tau_1\right] \tag{7}$$

with $\hat{T}$ being the time-ordering operator. If we consider the maximization of $J_I$, for example, the first-order variation of $J_I$ is given by

$$\delta J_I = \int_0^{t_f} dt \left\{ 2\operatorname{Re}\frac{i}{\hbar}\int_t^{t_f} d\tau\, y(\tau)\langle\langle W(\tau)|G(\tau,t)M|\rho(t)\rangle\rangle - \frac{2}{\hbar A(t)}E(t)\right\}\delta E(t), \tag{8}$$

where we have adjusted the integral parameters. If we further introduce the Liouville-space vector defined by

$$|\xi(t)\rangle\rangle = \int_t^{t_f} d\tau\, G^\dagger(\tau,t) M |W(\tau)\rangle\rangle y(\tau), \tag{9}$$

we will have the expression of the optimal pulse

$$E(t) = -A(t)\operatorname{Im}\langle\langle\xi(t)|M|\rho(t)\rangle\rangle. \tag{10}$$

It is easy to see that we are able to derive the same expression of the optimal pulse as that in Eq. (10) if we start with the unconstrained functional, in which the constraint due to Eq. (2) is explicitly represented by the Lagrange multiplier. In fact, the vector defined by Eq. (9) is the Lagrange multiplier that represents the constraint [29, 30]. It is easy to see that $|\xi(t)\rangle\rangle$ obeys the differential equation,

$$\frac{\partial}{\partial t}|\xi(t)\rangle\rangle = -\frac{i}{\hbar}L_{\text{eff}}^{\dagger}(t)|\xi(t)\rangle\rangle - y(t)|W(t)\rangle\rangle, \tag{11}$$

with the final condition $|\xi(t_{\text{f}})\rangle\rangle = 0$. When solving the coupled pulse-design equations, we use Eq. (11) instead of Eq. (9).

For the maximization of $J_{II}$, we have the same optimal pulse expression as that in Eq. (10). However, the equation of motion for the Lagrange multiplier is slightly different from Eq. (11). The Lagrange multiplier derived from the type II functional obeys a similar equation of motion to Eq. (11) but with the inhomogeneous term $y(t)|W(t)\rangle\rangle$ replaced by $|W(t)\rangle\rangle y(t)\langle\langle W(t)|\rho(t)\rangle\rangle$.

### III. Results and discussion

We assume the two-level system consisting of $|0\rangle$ and $|1\rangle$, the energy difference of which is given by the dimensionless value, $\omega_{10} = \omega_1 - \omega_0 = 30$. The system is assumed to be initially in the ground state $|0\rangle$. The transition matrix elements are set to $\langle 1|\mu|0\rangle = \langle 0|\mu|1\rangle = 1$ so that the magnitude of the dimensionless interaction represents the dimensionless optimal pulse. The relaxation operator $\Gamma$ is characterized by the matrix elements $\gamma^{(d)} = \langle\langle 10|\Gamma|10\rangle\rangle = \langle\langle 01|\Gamma|01\rangle\rangle$ and $\gamma_{0\leftarrow 1} = \langle\langle 00|\Gamma|11\rangle\rangle$ associated with the dephasing and the population decay, respectively. In the numerical applications, we assume the rotating-wave approximation (RWA), in which the counter-rotating components are neglected. The dimensionless final time is set to $t_{\text{f}} = 25$, which is uniformly divided into 30000 grid points. The time evolution is calculated by using the 4$^{\text{th}}$-order Runge-Kutta method.

For the type I functional in Eq. (3), we adopt the Hermitian target operator

$$W(t) = \frac{1}{2}\left(|1\rangle e^{-i\omega t - i\theta}\langle 0| + |0\rangle e^{i\omega t + i\theta}\langle 1|\right), \tag{12}$$

where $\omega$ and $\theta$ are the target frequency and the phase shift, respectively. Here, we assume $\omega = \omega_{10}$ for simplicity. Although we assume $\theta = 0$ in the following, we have checked that the optimal pulse changes its phase according to the value of $\theta$ (not shown here). The envelope function $y(t)$ that specifies the control period is expressed as the sum of the following function

$$y(t) = \sum_{j=1} (-1)^{j-1} \frac{1}{1+e^{-\alpha(t-t_j)}}, \qquad (13)$$

where $0 < t_1 < t_2 \cdots$. The parameter $\alpha = 3.0$ is assumed to avoid the abrupt change of $y(t)$. For the penalty term in Eq. (5), the positive function $A(t)$ is expressed as

$$A(t) = A_0 \times \begin{cases} \sin(\frac{t}{2T}\pi) & 0 \leq t < T \\ 1 & T \leq t \leq t_f - T, \\ \sin(\frac{t_f - t}{2T}\pi) & t_f - T < t \leq t_f \end{cases} \qquad (14)$$

where the parameter $T = 1.0$ prevents the abrupt rise and fall of the optimal pulse at the initial time and the final time. The parameter $A_0$ determines the magnitude of the penalty. The smaller (larger) value of $A_0$ monotonically leads to the lower (higher) pulse fluence, which is numerically illustrated in Appendix A. We actively utilize this property to design the optimal pulse with the specified pulse fluence $f_0$ by adjusting the value of $A_0$. In Appendix A, the convergence behavior as a function of the iteration is shown for $\gamma^{(d)} = 0$ and $\gamma^{(d)} = 0.3$ when $f_0 = 0.2$ is assumed. We see that the fluence-specified optimal control simulation is realized regardless of the maximization problem and of the absence or presence of decoherence.

Figure 1 shows the optimally designed external pulses as a function of time for several values of $\gamma^{(d)}$ when the pulse fluence is set to 0.2. The control period in Eq. (13) with $t_1 = 5$ and $t_2 = 20$ is illustrated in Fig. 1(c), which leads to the "ideal" value of $(F_I)_{\text{ideal}} = 15$. In the limiting case of no dephasing ($\gamma^{(d)} = 0$), the optimal pulse in Fig. 1(a) creates the target superposition state before the control period starts ($t_1 = 5$) and achieves 99.9% of the ideal value. As the value of $\gamma^{(d)}$ increases, the temporal peaks around $t_1 = 5$ of the optimal pulses appear to quickly excite the population against the dephasing (Figs. 1(b)–(e)). In addition, temporally broad components gradually broaden over the control period as $\gamma^{(d)}$ increases. The broad components successively and gradually create the target superposition state while partially suppressing the dephasing. The above interpretations can be confirmed by the time evolution of the populations and that of the integrand of $F_I$, i.e., $2\text{Tr}\{W(t)\rho(t)\}$ in Fig. 2. When $\gamma^{(d)} \leq 0.10$, the two states are almost equally populated after the control. On the other hand, when $\gamma^{(d)} \geq 0.20$, the population of the excited state $|1\rangle$ is below 0.5 even at the final time because the dephasing removes the system coherence much more quickly than the excitation processes. We see from Fig. 2(e) that $2\text{Tr}\{W(t)\rho(t)\}$ has a flat structure and evolves in time almost parallel to the envelope function $y(t)$.

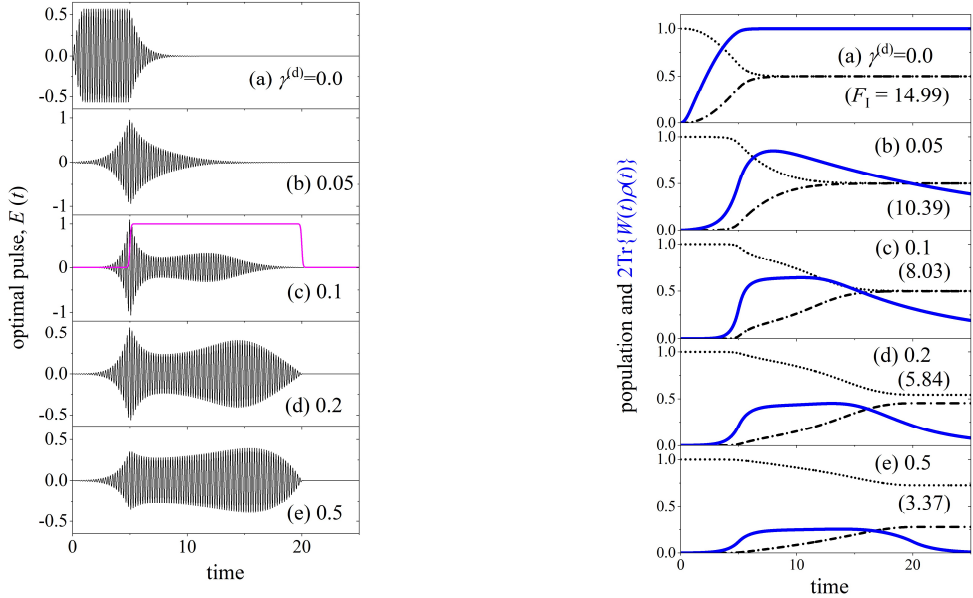

Figure 1 (Left)
Optimal pulses designed by using the type I functional (maximization of $J_I$) in the cases of (a) $\gamma^{(d)} = 0.0$, (b) 0.05, (c) 0.1, (d) 0.2, and (e) 0.5. The purple solid line in (c) shows the control period, i.e., $y(t)$ in Eq. (13) with $t_1 = 5$ and $t_2 = 20$.

Figure 2 (Right)
Time evolution of the integrand of $F_I$, i.e., $2\text{Tr}\{W(t)\rho(t)\}$ (solid lines), and that of the populations of the states, $|0\rangle$ (dotted lines) and $|1\rangle$ (dot-dashed lines), when the target operator is given by Eq. (12). Values in parentheses show the target expectation values $F_I$'s. The results in (a)–(e) are obtained by using the optimal pulses shown in Figs. 1(a)–(e), respectively.

Figure 3 summarizes the target expectation value $F_I$ as a function of $\gamma^{(d)}$ for several specified values of the pulse fluence. The control period is the same as that used in Fig. 1. As expected, for a given value of the pulse fluence, $F_I$ monotonically decreases as the value of $\gamma^{(d)}$ increases. For a given value of $\gamma^{(d)}$, $F_I$ increases as the pulse fluence increases. Note that when $\gamma^{(d)} \leq 0.1$, the optimal pulses whose fluence is higher than $f_0 = 0.1$ yield almost the same target expectation values. The target expectation values associated with $f_0 = 0.2$ are similar to those associated with $f_0 = 0.3$ in the entire range of $\gamma^{(d)}$ in Fig. 3. The above results suggest that there is a "saturated" value with respect to the pulse fluence for a given value of $\gamma^{(d)}$. To confirm this "saturation", we optimize the control pulse without the penalty (penalty-free optimization) in the case of $\gamma^{(d)} = 0.5$ as an example and show the result as a solid star in Fig. 3. The saturation may originate from the Markovian approximation, in which the system coherence spread over the environment quickly vanishes and does not return to the system.

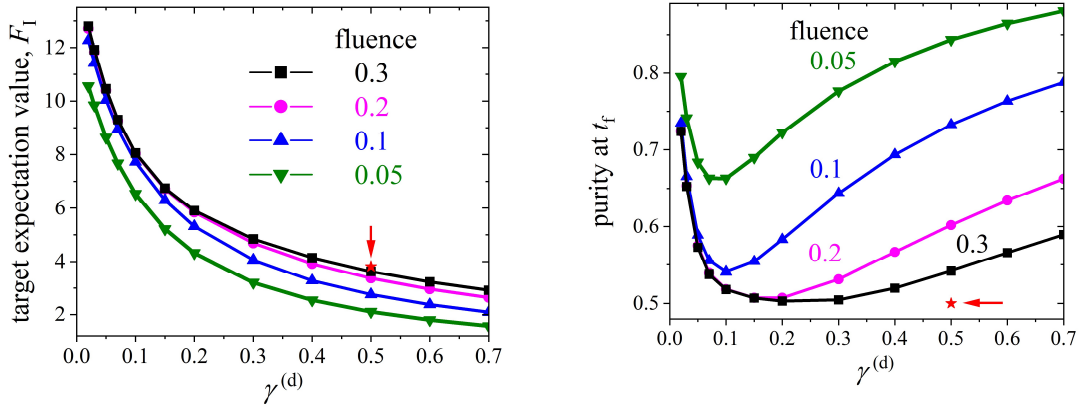

Figure 3 (Left)
Target expectation value $F_I$ as a function of $\gamma^{(d)}$ for several specified values of the pulse fluence. The solid star indicated by an arrow shows the result obtained by the penalty-free optimization, i.e., $J_I = F_I$, in which the convergence criterion, $J_I^{(k)} - J_I^{(k-1)} \leq 10^{-8}$ with $k$ being the iteration steps, is assumed.

Figure 4 (Right)
Purity at the final time, i.e., $\mathrm{Tr}\{\rho^2(t_f)\}$ as a function of $\gamma^{(d)}$ for several specified values of the pulse fluence, which are the same as those used in Fig. 3. The solid star shows the result obtained by the penalty-free optimization (see the caption of Fig. 3).

From the results in Figs. 1 to 3, it is natural to ask how much the optimal pulses actively utilize the system coherence, which is measured by the purity defined by the trace of $\rho^2(t)$, i.e., $\mathrm{Tr}\{\rho^2(t)\}$ [31-33]. The purity of a pure state has a value of 1. If the system is completely decohered by the dephasing and is reduced to the mixed state that corresponds to the maximum entropy, then the value of purity is 0.5. Figure 4 shows the purity at the final time $t_f = 25$ as a function of $\gamma^{(d)}$, where each line corresponds to the result in Fig. 3. From the downward convex structures of the lines, it is convenient to consider the control mechanisms separately according to the values of $\gamma^{(d)}$. For small values of $\gamma^{(d)} \leq 0.05$, the system coherence can survive at $t_f$ to some extent and therefore, even the optimal pulses with $f_0 = 0.1$ can realize the saturated target expectation values. For intermediate values of $0.1 \leq \gamma^{(d)} \leq 0.2$, a high pulse fluence is required to achieve the saturated values by fully utilizing the system coherence, which leads to the minimum value of purity 0.5. When $\gamma^{(d)} \geq 0.3$, the values of $\gamma^{(d)}$ are too large to effectively create the target superposition state, thereby resulting in the increase in the "residual" purity at $t_f$. In fact, the optimal pulse derived by the penalty-free optimization leads to $\mathrm{Tr}\{\rho^2(t_f)\} = 0.5$ (see the solid star in Fig. 4); however, it gives rise to a quite small improvement of the target expectation value, as shown in Fig. 3.

We next consider the effects of the population decay, which can reduce the entropy, i.e., increase the purity of the system. Figure 5 shows the target expectation value as a function of $\gamma_{0\leftarrow 1}$ when $f_0 = 0.2$ and $\gamma^{(d)} = 0.2$ are assumed. The increase in value of $\gamma_{0\leftarrow 1}$ monotonically increases the target expectation value $F_I$. As an example, let us consider the case of $\gamma_{0\leftarrow 1} = 0.1$ and compare the results in Fig. 6 with those calculated by the optimal pulses in Fig. 1(d) ($\gamma_{0\leftarrow 1} = 0$). Although the population decay does not significantly improve the value of $F_I$ (Fig. 5), the control mechanisms are significantly different from those in the absence of the population decay. The optimal pulse in Fig. 6(a) has a slightly flatter shape and starts the excitation much earlier than that in Fig. 1(d). The superposition state created earlier may face a greater possibility of being destroyed by the dephasing but at the same time have a greater possibility of being purified by the population decay. The optimal pulse in Fig. 6(a) finds the balance between the trade-off and excites a much smaller population in Fig. 6(b) than in Fig. 2(d), leading to a slightly larger value of $2\mathrm{Tr}\{W(t)\rho(t)\}$ than that in Fig. 2(d). From a qualitative viewpoint, we may consider the time derivative $d\mathrm{Tr}\{\rho^2(t)\}/dt = 2\mathrm{Tr}\{\rho(t)\dot\rho(t)\}$ according to Ref. [32], which measures the temporal change in the system's purity. In the present model, it is easy to derive

$$\frac{d}{dt}\text{Tr}\{\rho^2(t)\} = 2\gamma_{0\leftarrow 1}\{2\rho_{00}(t)-1\}\{1-\rho_{00}(t)\} - 4\gamma^{(d)}|\rho_{01}(t)|^2, \tag{15}$$

where $\rho_{mn}(t) = \langle m|\rho(t)|n\rangle$. We see from Eq. (15) that the population decay can slow down the decrease in purity if $\{2\rho_{00}(t)-1\}\{1-\rho_{00}(t)\} > 0$ that has the maximum value when $\rho_{00}(t) = 0.75$. This could explain the reason why $\rho_{00}(t)$, i.e., the population of the state $|0\rangle$ is almost constant ~0.75 during the control period in Fig. 6(b).

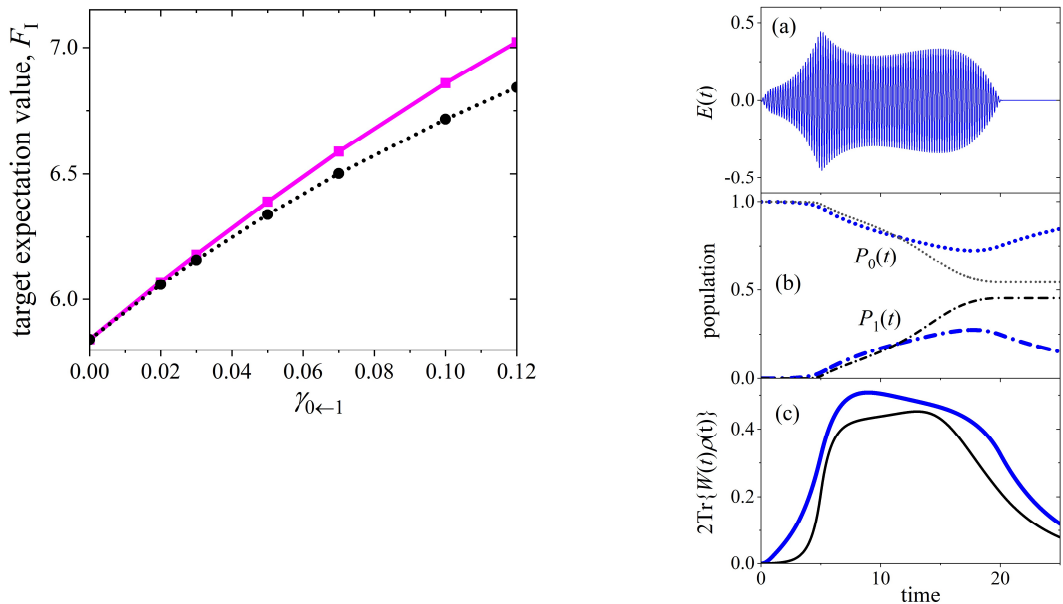

Figure 5 (Left)
Target expectation value $F_I$ as a function of $\gamma_{0\leftarrow 1}$ when $\gamma^{(d)} = 0.2$ and $f_0 = 0.2$ are assumed. For reference, the dotted line shows the target expectation values calculated by using the optimal pulse designed without $\gamma_{0\leftarrow 1}$ [Fig. 1(d)].

Figure 6 (Right)
(a) Optimal pulse as a function of time derived by the type I functional (maximization of $J_I$) by including the population decay, in which $f_0 = 0.2$, $\gamma^{(d)} = 0.2$, and $\gamma_{0\leftarrow 1} = 0.1$ are assumed. (b) Time evolution of the populations of states $|0\rangle$ and $|1\rangle$ is shown by the blue dotted and dot-dashed lines, respectively, and (c) that of the integrand of $F_I$ is shown by the blue solid line. In (b) and (c), for reference, the time evolution of those driven by the optimal pulse in Fig. 1(d) is also plotted by the thin lines.

Finally, we consider the optimal pulses derived by the maximization of $J_{II}$. Referring to the type I target operator in Eq. (12), we adopt the following non-Hermitian operator

$$W(t) = 2|1\rangle\langle 0|, \tag{16}$$

where the control period $y(t)$ is set to the same value as that used in Fig. 1. Because of the constant 2 in Eq. (16), the type II functional $F_{II}$ in Eq. (4) has the same ideal value of 15 as the type I functional. In Eq. (16), there are no (time-dependent) phase factors because of the absolute square of the expectation value so that the integrand of the type II functional does not depend on the phase factors.

We solve the coupled-pulse design equations derived from the maximization problem of $J_{II}$. Figure 7 shows the optimal pulses for several values of $\gamma^{(d)}$. The parameters used in the simulation are the same as those in Fig. 1. In the case of small dephasing $\gamma^{(d)} \leq 0.05$, the optimal pulses in Figs. 1(a) and 7(a) and those in Figs. 1(b) and 7(b) have almost the same shapes.

As the value of $\gamma^{(d)}$ increases, the optimal pulses in Figs. 7(d) and 7(e) tend to have broad, single-peak structures, whereas those in Figs. 1(d) and 1(e) are characterized by double-peak structures. This clearly indicates that the solution (optimal pulse) is strongly dependent on the functional forms when the decoherence is not negligible.

Before discussing the above-mentioned difference, we show the time evolution of the populations in Fig. 8, from which we see that the temporal behavior is similar to that in Fig. 2. In Fig. 8, we also show the time evolution of the absolute value of the expectation value $|\text{Tr}\{W(t)\rho(t)\}|$ (not the integrand of the type II functional) and conveniently compare it to the temporal behavior of the integrand of the type I functional (Fig. 2). When the values of $\gamma^{(d)}$ increase, the expectation values show different time-dependent behaviors, which can be expected from the different shapes of the optimal pulses in Figs. 1 and 7. In the case of $\gamma^{(d)} = 0.5$, for example, the expectation value in Fig. 2(e) has a flat structure, whereas that in Fig. 8(e) has a bell-shaped structure.

We now consider the difference between the optimal pulses in Figs. 1 and 7. To confirm that there is no significant difference between them, we substitute the optimal pulses in

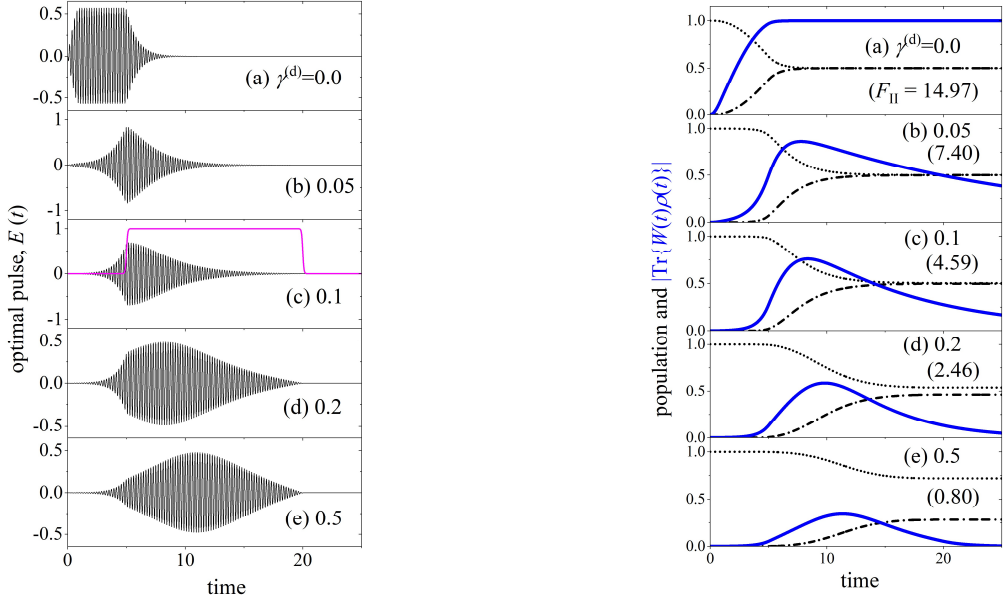

Figure 7 (Left)
Optimal pulses designed by using the type II functional (maximization of $J_{II}$) in cases of (a) $\gamma^{(d)}=0.0$, (b) 0.05, (c) 0.1, (d) 0.2, and (e) 0.5. The purple solid line in (c) shows the control period, i.e., $y(t)$ in Eq. (13) with $t_1=5$ and $t_2=20$.

Figure 8 (Right)
Time evolution of the absolute value of the expectation value $|\text{Tr}\{W(t)\rho(t)\}|$ (solid lines), and that of the populations of states $|0\rangle$ (dotted lines) and $|1\rangle$ (dot-dashed lines), when the non-Hermitian target operator in Eq. (16) is assumed. Note that the integrand of the type II functional is given by the square of the absolute value. Values in parentheses show the target expectation values $F_{II}$'s. The results in (a)–(e) are obtained by using the optimal pulses in Figs. 7(a)–(e), respectively.

Figs. 7(a)–(e) into the type I functional $F_I$ and obtain the values of (a) 14.99 (100%), (b) 10.39 (100%), (c) 7.95 (99%), (d) 5.61 (96%), and (e) 3.18 (94%). Here, the values divided by the original values in Fig. 2 are given by the ratios (%) in parentheses. The optimal pulses derived from $J_{II}$ can reproduce the values of $F_I$ with high probability even in the cases of the large dephasing. Qualitatively, this means that the common features of the temporally broadened structures of the optimal pulses are most important to maximally create and maintain the present target superposition state. In spite of this, we may say that if we need to design an optimal pulse that is highly tuned to a specified physical objective and/or that controls the intermediate dynamics during the control period, it will be essential to adopt a suitable functional that is adjusted to the control objective.

Finally, we examine the controlled dynamics graphically. The Bloch vector may be the first choice for this purpose, in which the dynamics is described by the trajectory of the three-dimensional vector defined by $\boldsymbol{R}(t)=\text{Tr}\{\rho(t)\boldsymbol{\sigma}\}$ with $\boldsymbol{\sigma}=(\sigma_x,\sigma_y,\sigma_z)$ being the Pauli matrices. In the present study, on the other hand, we are primarily interested in the magnitude of $\langle 0|\rho(t)|1\rangle$, not in the full information of $\langle 0|\rho(t)|1\rangle$. If we focus on the absolute value $|\langle 0|\rho(t)|1\rangle|$, there may be another way of graphical analysis. On the basis of the previous study [32], we consider the purity that can be expressed in terms of the matrix elements of the density operator such that

$$\mathrm{Tr}\{\rho^2(t)\} = 2\rho_{00}^2(t) - 2\rho_{00}(t) + |\rho_{01}(t)|^2 + 1, \qquad (17)$$

where $\rho_{mn}(t) = \langle m|\rho(t)|n\rangle$. Figure 9 shows the trajectories on the contour map of the purity as a function of $\rho_{00}$ and $|\rho_{01}|$, which is the modified version of Ref. [32]. Here, we have derived the trajectory by regarding $\rho_{00}(t)$ and $|\rho_{01}(t)|$ as the parametric representation of a curve so that we have removed "$(t)$" in Fig. 9. The trajectories correspond to the results in Fig. 1 (solid lines) and Fig. 7 (dashed lines). The temporal behavior is indicated by the solid circles and triangles that are plotted at every 2.5 from $t = 0$ to $t_f = 25$. Here, the trajectory derived from the maximization of $J_I$ ($J_{II}$) is referred to as the type I (type II) trajectory, for convenience.

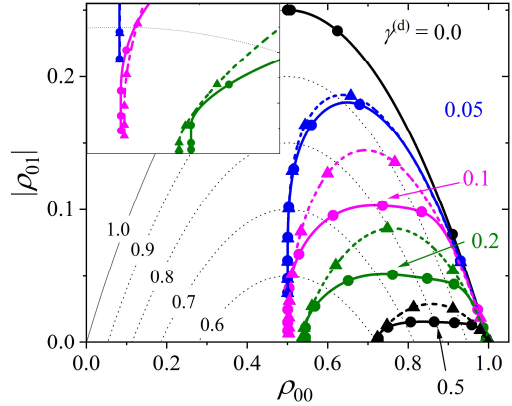

Figure 9
Trajectories on the contour map of the purity as a function of $\rho_{00}$ and $|\rho_{01}|$ for $\gamma^{(d)} = 0.0$ (black), 0.05 (blue), 0.1 (purple), (d) 0.2 (green), and (e) 0.5 (black). Solid (dashed) lines show the results derived by the type I (II) functional. Solid circles (triangles) indicate the time evolution and appear every 2.5 from $t = 0$ to $t_f = 25$. The contours are specified by the thin dashed lines, the values of which are shown. The inset is the enlarged figure around $\rho_{00} = 0.5$ and $|\rho_{01}| = 0.6$.

We see from Fig. 9 that the differences between the types I and II trajectory become more noticeable as the value of $\gamma^{(d)}$ increases. For a given value of $\gamma^{(d)}$, on the other hand, the types I and II trajectory always meet at almost the same point at the final time (see also the inset of Fig. 9). This means that the optimal pulses designed by using the types I and II functional utilize almost the same amount of system coherence. In this sense, we may say that there is no superiority or inferiority between the types I and II functional for the present control objective. Focusing on the trajectories during the control period, however, we also see a different temporal behavior between the types I and II trajectory particularly when $\gamma^{(d)}$ has an "intermediate" value, i.e., $\gamma^{(d)} \sim 0.1$ in Fig. 9. In such a case, it is necessary to choose a functional to better quantify the control objective associated with the intermediate dynamics. In the case of a large value of $\gamma^{(d)}$, i.e., $\gamma^{(d)} \sim 0.5$ in Fig. 9, it is difficult to effectively create and/or maintain the target superposition state. Because of this, the types I and II trajectory are restricted to a small area so that the difference between them becomes less noticeable.

**Summary**

We have numerically studied how to maximally create and maintain a specified superposition state of a two-level system during a specified control period under the influence of the Markovian decoherence. We numerically designed the control pulses by using the optimal control theory with types I and II functional subject to the penalty due to pulse fluence. The type I (II) functional is given by the integral over the specified control period, the integrand of which is expressed in terms of the (absolute square of) off-diagonal elements of the density operator. We have simulated the optimal pulses while systematically varying the values of the dephasing $\gamma^{(d)}$, the population decay $\gamma_{0\leftarrow 1}$, the pulse fluence $f_0$, and the control period. We have also proposed graphical analyses to examine the control mechanisms by using the trajectories on the $\rho_{00}$-$|\rho_{01}|$ contour map.

The numerical findings are summarized as follows. The decrease in system's purity due to the Markovian decoherence imposes the restriction on the realization of the specified superposition state, which results in the saturation of the target expectation value with respect to the pulse fluence. As the value of $\gamma^{(d)}$ increases, the temporarily broadened optimal pulses effectively deal with the decoherence by gradually creating the target superposition state to achieve the saturated target expectation value as much as possible. The presence of the population decay can partly recover the system coherence and slow down the speed of the decrease in purity, which considerably change the control mechanisms. Qualitatively, the above-mentioned numerical features are common independent of the types of the functional.

In the present study, we have restricted ourselves to the initial-state-dependent DD under the Markovian decoherence according to the widely used CPMG technique. The straightforward extension to the initial-state-independent DD under the non-Markovian decoherence [34] will be discussed elsewhere.

**Acknowledgements**

It is with deep gratitude that we dedicate this study to the memory of Professor Sheng Hsien Lin. One of the authors (YO) acknowledges support from a Grant-in-Aid for Scientific Research (C) (20K05414) and partial support from the Joint Usage/Research Program on Zero-Emission Energy Research, Institute of Advanced Energy, Kyoto University (ZE2022B-05).

**APPENDIX A: Fluence-specified optimal control simulation**

We briefly summarize the numerical techniques for designing an optimal pulse with a specified pulse fluence. We obtain the optimal pulses by adopting several values of $A_0$ [Eq. (14)] and plot the target expectation value as a function of the pulse fluence, as shown in Fig. 10 for several values of $\gamma^{(d)}$. It is clear that the larger value of $A_0$ monotonically leads to the larger target expectation value and the higher pulse fluence. This suggests that we could specify the value of the pulse fluence by suitably adjusting the value of $A_0$.

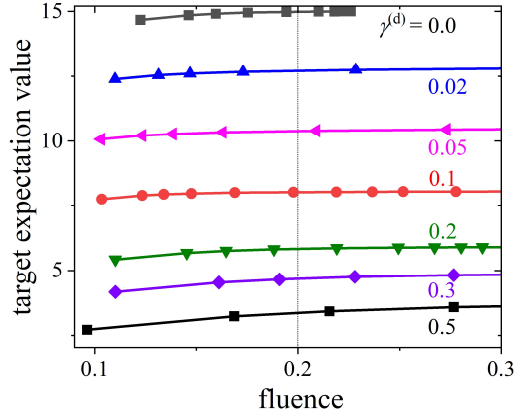

Figure 10
Target expectation value $F_I$ as a function of pulse fluence in the range of $[0.09, 0.3]$ for several values of $\gamma^{(d)}$. Solid squares, triangles, etc. show $F_I$ obtained by using the amplitude parameters $A_0 = 0.1, 0.2, 0.3, 0.5, 1.0, 2.0, 3.0, 4.0,$ and $5.0$ [see Eq. (14)] from left to right. For convenience, the vertical dotted line shows the pulse fluence $f_0 = 0.2$.

To realize the pulse-fluence-specified simulation, we introduce the iteration-step-dependent amplitude parameters and use the following procedure to specify the fluence $f_0$. As an example, we consider the $k$th iteration step, in which the amplitude parameter and the pulse fluence are given by $A_0^{(k)}$ and $f^{(k)}$, respectively. By using these values, the amplitude parameter in the next step is empirically determined by $A_0^{(k+1)} = \sqrt{f_0/f^{(k)}} A_0^{(k)}$. To smoothly adjust the value of $A_0^{(k+1)}$, the maximum change of value is set to 0.1%. We introduce two extra criteria associated with the pulse fluence

$$\delta_0^{(k)} = (f^{(k)} - f_0)/f_0 \quad \text{and} \quad \delta_1^{(k)} = (f^{(k)} - f^{(k-1)})/f^{(k)} \tag{A1}$$

to determine the convergence, that is, we require $\delta_0^{(k)} \leq 10^{-4}$ and $\delta_1^{(k)} \leq 10^{-6}$, the values of which are also determined empirically. We also require the convergence criterion of the functional $0 < F_I^{(k+1)} - F_I^{(k)} \leq 10^{-8}$. If all the convergence criteria are satisfied simultaneously and successively 30 times, the iteration is considered converged. We assume the same convergence criteria when calculating the optimal pulses with the type II functional. In the present study, the value of the pulse fluence after the convergence coincides with $f_0$ by at least 10 digits (typically more) if we start the iteration from a suitable initial guess pulse. Figure 11 shows the typical examples of the convergence behaviors in the case of $f_0 = 0.2$. We see good convergence behavior that is independent of the type of functional and of the absence/presence of dephasing.

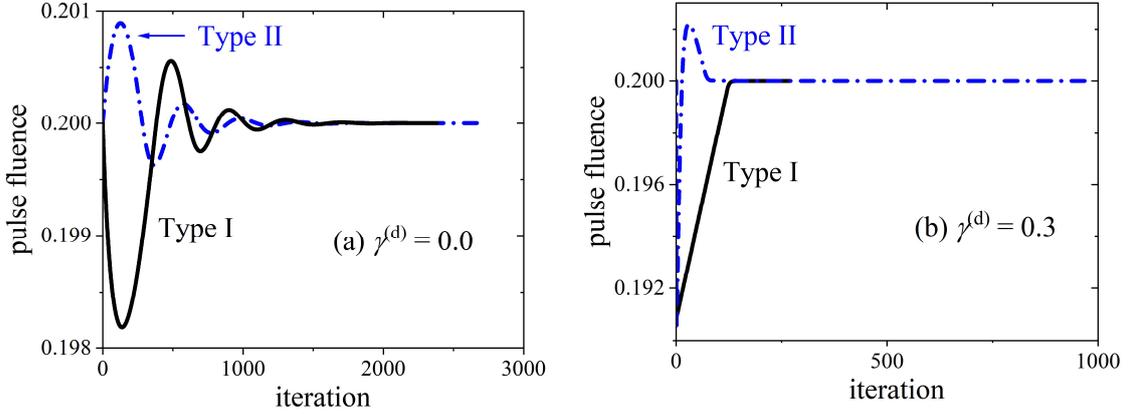

Figure 11
Convergence behavior of the fluence-specified optimal control simulation in the case of (a) $\gamma^{(d)} = 0.0$ and (b) $\gamma^{(d)} = 0.3$ when the target pulse fluence is set to 0.2. The results associated with the types I and II functional are plotted by solid and dot-dashed lines, respectively.

The results in Fig. 10 suggest that a larger (smaller) value of the amplitude parameter seems to lead to larger (smaller) values of the objective functional, the target expectation as well as the pulse fluence. In fact, the above-mentioned amplitude-parameter dependence can be proved in more general cases. For this purpose, we consider a general maximization problem defined by

$$J = K - \frac{1}{\lambda} \int_0^{t_f} dt \frac{[E(t)]^2}{\eta(t)} \equiv K - \frac{f}{\lambda}, \tag{A2}$$

where $K$ represents all the terms associated with the target expectation value. In Eq. (A2), the positive function that weighs the physical significance of the penalty is divided into the time-independent and time-dependent parts ($\lambda$ and $\eta(t)$) for convenience. The pulse fluence integrated with the weight function $\eta(t)$ is given by $f$ in Eq. (A2). If we consider the case where the optimal pulse $E_A(t)$ [$E_B(t)$] maximizes the functional when $\lambda = \lambda_A$ ($\lambda = \lambda_B$), then we will have the maximized objective functionals

$$J_A = K_A - \frac{f_A}{\lambda_A} \quad \text{with} \quad f_A = \int_0^{t_f} dt \frac{[E_A(t)]^2}{\eta(t)} \tag{A3}$$

and

$$J_B = K_B - \frac{f_B}{\lambda_B} \quad \text{with} \quad f_B = \int_0^{t_f} dt \frac{[E_B(t)]^2}{\eta(t)}, \tag{A4}$$

which are derived by the optimal pulses $E_A(t)$ and $E_B(t)$, respectively. In the following, we assume that $\lambda_A > \lambda_B$ without loss of generality.

In the next step, we replace $\lambda_B$ by $\lambda_A$ in Eq. (A4) [$\lambda_A$ by $\lambda_B$ in Eq. (A3)] to obtain

$$\tilde{J}_A = K_B - \frac{f_B}{\lambda_A} \quad \text{and} \quad \tilde{J}_B = K_A - \frac{f_A}{\lambda_B}. \tag{A5}$$

Because the optimal pulses $E_A(t)$ and $E_B(t)$ are the solutions of the maximization problems, $J_A > \tilde{J}_A$ and $J_B > \tilde{J}_B$ should be justified. From the two inequalities, we obtain

$$\frac{f_A - f_B}{\lambda_B} > K_A - K_B > \frac{f_A - f_B}{\lambda_A}, \tag{A6}$$

which leads to $J_A > J_B$, $K_A > K_B$ and $f_A > f_B$ because of $\lambda_A > \lambda_B$.

## APPENDIX B: Optimal control simulation with two control periods

We numerically optimize the pulses assuming the control period, which consists of the two parts specified by $y(t)$ in Eq. (13) with $t_1 = 5$, $t_2 = 10$, $t_3 = 15$, and $t_4 = 20$. We adopt several values of the pulse fluence while assuming the fixed value $\gamma^{(d)} = 0.5$ and show the results in Figs. 12 and 13. In Fig. 12(c), the control period is shown by the purple line. For convenience, the first (second) part of the control period is referred to as the first (second) period. Roughly speaking, the optimal pulses in Fig. 12 are composed of two pulses corresponding the first and second periods, the fluence ratios of which are given by (a) 1:1.78, (b) 1:2.23, (c) 1:2.69, (d) 1:2.54, and (e) 1:2.20. Although the pulses associated with the first period have approximately half smaller values of fluence than those associated with the second period, the ratios of the target expectation values are almost 1:1, i.e., (a) 1:1.04, (b) 1:1.06, (c) 1:1.04, (d) 1:0.99, and (e) 1:0.96. Although the increase in pulse fluence leads to more structured optimal pulses, the target expectation values converge to a "saturated" value, as shown in Fig. 13.

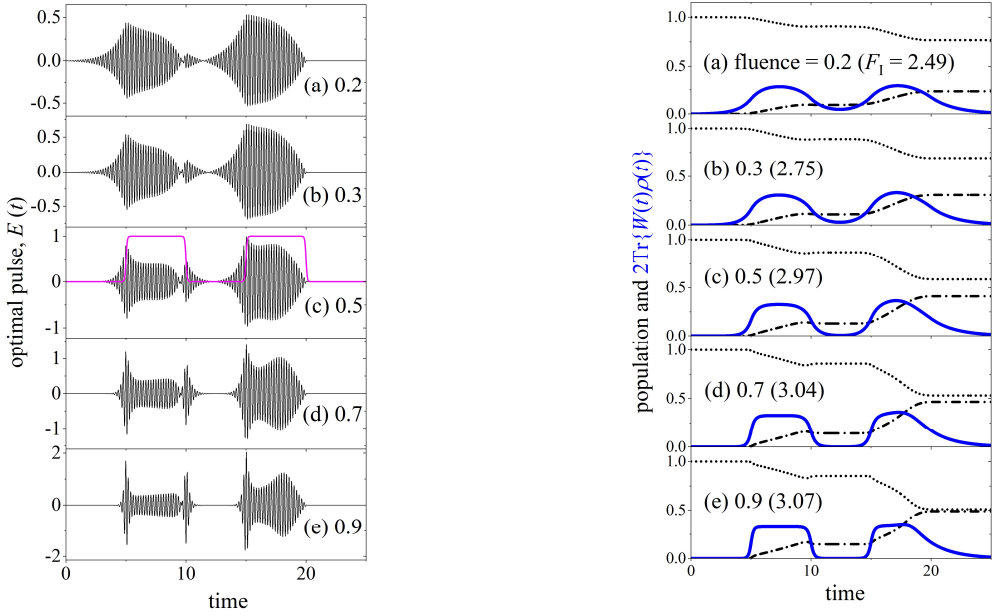

Figure 12 (Left)
Optimal pulses designed by using the type I functional (maximization of $J_I$) in the case $\gamma^{(d)} = 0.5$ for several values of pulse fluence, (a) 0.2, (b) 0.3, (c) 0.5, (d) 0.7, and (e) 0.9. The purple solid line in (c) shows the control period, i.e., $y(t)$ in Eq. (13) with $t_1 = 5$, $t_2 = 10$, $t_3 = 15$, and $t_4 = 20$.

Figure 13 (Right)
Time evolution of the integrand of $F_I$, i.e., $2\text{Tr}\{W(t)\rho(t)\}$ (solid lines), that of the populations of states $|0\rangle$ (dotted lines) and $|1\rangle$ (dot-dashed lines) when the target operator is given by Eq. (12). Values in parentheses show the target expectation values $F_I$'s. The results in (a)–(e) are obtained by using the optimal pulses shown in Figs. 12(a)–(e), respectively.

optimally designed gate pulses, *New. J. Phys.* **12**, 045002 (2010).